\documentclass[a4paper,12pt, epsfig]{article}
\usepackage{epsfig}
\usepackage{amssymb}
\usepackage{amsfonts}

\newskip\humongous \humongous=0pt plus 1000pt minus 1000pt

\newif\ifdtup

\catcode`\@=11

\@addtoreset{equation}{section}
\def\theequation{\thesection.\arabic{equation}}

\def\@normalsize{\@setsize\normalsize{15pt}\xiipt\@xiipt
\abovedisplayskip 14pt plus3pt minus3pt%
\belowdisplayskip \abovedisplayskip
\abovedisplayshortskip \z@ plus3pt%
\belowdisplayshortskip 7pt plus3.5pt minus0pt}

\def\small{\@setsize\small{13.6pt}\xipt\@xipt
\abovedisplayskip 13pt plus3pt minus3pt%
\belowdisplayskip \abovedisplayskip
\abovedisplayshortskip \z@ plus3pt%
\belowdisplayshortskip 7pt plus3.5pt minus0pt
\def\@listi{\parsep 4.5pt plus 2pt minus 1pt
     \itemsep \parsep
     \topsep 9pt plus 3pt minus 3pt}}

\relax

\catcode`@=12

\catcode`\@=11

\def\section{\@startsection{section}{1}{\z@}{3.5ex plus 1ex minus
   .2ex}{2.3ex plus .2ex}{\large\bf}}

\def\thesection{\arabic{section}}
\def\thesubsection{\arabic{section}.\arabic{subsection}}

\def\appendix{\setcounter{section}{0}
\def\thesection{Appendix \Alph{section}}
\def\thesubsection{\Alph{section}.\arabic{subsection}}
\def\theequation{\Alph{section}.\arabic{equation}}}

\def\YGrule{0.4}   
\def\YGbox{6.5}    
\def\SymBoxes#1#2#3#4{\newdimen\un@t \un@t#3%
\raisebox{#1}{\rule{#2\un@t}{#4}\hskip-#2\un@t
\@tempdimb\un@t \advance\@tempdimb by-#4\@tempcntb#2\relax%
\@whilenum{\@tempcntb>0}\do{
\rule{#4}{\un@t}\hskip\@tempdimb \advance\@tempcntb by\m@ne}%
\hskip-#2\un@t \rule[\un@t]{#2\un@t}{#4}%
\rule[\un@t]{#4}{#4}\hskip-#4
\rule{#4}{\un@t}}\hskip-#4}                
\def\Young{\@ifnextchar[{\@Young}{\@Young[0]}}
\def\@Young[#1]#2{\newdimen\YG@unit \YG@unit\YGbox pt%
\newdimen\h@ight \h@ight#1\YG@unit \@tempcnta-1\relax
\@tfor\c@ount:=#2\do{\advance\@tempcnta by\@ne}
\@tempdima\@tempcnta\YG@unit%
\advance\h@ight by\@tempdima\relax     
\@tfor\c@ount:=#2\do{\SymBoxes{\h@ight}{\c@ount}{\YG@unit}{\YGrule pt}%
\@tempdima-\c@ount\YG@unit \hskip\@tempdima%
\advance \h@ight by -\YG@unit}         
\@tempdima\YG@unit \multiply\@tempdima by\@car#2\@nil %
\hskip\@tempdima}                      
\def\YoungTab{\@ifnextchar[{\@YoungIdx}{\@YoungIdx[0]}}
\def\@YoungIdx[#1]{\@ifnextchar[{\@iYoungIdx[#1]}{\@iYoungIdx[#1][\@empty]}}
\def\@iYoungIdx[#1][#2]#3{%
\newdimen\YG@unit \YG@unit\YGbox pt\newdimen\YG@rule \YG@rule \YGrule pt
\newcount\c@ount \c@ount\z@ \newdimen\skip@wd \unitlength\@ne pt
\newdimen\h@ight \h@ight#1\YG@unit \@tempcnta\m@ne\relax
\@tfor\d@um:=#3\do{\advance\@tempcnta by\@ne}
\@tempdima\@tempcnta\YG@unit%
\advance\h@ight by\@tempdima\relax
\@tfor\@idxlist:=#3\do{
\@tempcnta\z@\hskip.5\YG@rule\relax
\@for\@idx:=\@idxlist\do{
\raisebox{\h@ight}{\makebox(\YGbox,\YGbox){#2$\@idx$}}
\advance\@tempcnta by\@ne}\hskip-.5\YG@rule%
\@tempdima-\@tempcnta\YG@unit \hskip\@tempdima%
\ifnum\c@ount=\z@ \skip@wd-\@tempdima\fi \relax
\SymBoxes{\h@ight}{\@tempcnta}{\YG@unit}{\YG@rule}%
\hskip\@tempdima \advance\h@ight by -\YG@unit
\advance\c@ount by\@ne}
\hskip\skip@wd}                      

\begin{document}

\newcommand{\be}{\begin{equation}}
\newcommand{\ee}{\end{equation}}
\newcommand{\bea}{\begin{eqnarray}}
\newcommand{\eea}{\end{eqnarray}}
\newcommand{\beas}{\begin{eqnarray*}}
\newcommand{\eeas}{\end{eqnarray*}}
\newcommand{\defi}{\stackrel{\rm def}{=}}
\newcommand{\non}{\nonumber}
\newcommand{\bquo}{\begin{quote}}
\newcommand{\enqu}{\end{quote}}
\def\de{\partial}
\def\Tr{ \hbox{\rm Tr}}
\def\const{\hbox {\rm const.}}
\def\o{\over}
\def\im{\hbox{\rm Im}}
\def\re{\hbox{\rm Re}}
\def\bra{\langle}\def\ket{\rangle}
\def\Arg{\hbox {\rm Arg}}
\def\Re{\hbox {\rm Re}}
\def\Im{\hbox {\rm Im}}
\def\diag{\hbox{\rm diag}}
\def\longvert{{\rule[-2mm]{0.1mm}{7mm}}\,}
\def\a{\alpha}
\def\dag{{}^{\dagger}}
\def\tq{{\widetilde q}}
\def\p{{}^{\prime}}
\def\W{W}
\def\N{{\cal N}}
\newcommand{\Z}{\ensuremath{\mathbb Z}}
\begin{titlepage}
\begin{flushright}
hep-th/0412239\\
\end{flushright}
\bigskip
\def\thefootnote{\fnsymbol{footnote}}

\begin{center}
{\large  {\bf
Volume Stabilization via $\alpha^\prime$ Corrections in Type IIB Theory with Fluxes
} }
\end{center}

\bigskip
\begin{center}
{\large  Konstantin Bobkov \footnote{\texttt{currently at DUMC: k.v.bobkov@duke.edu}}
\vskip 0.10cm
}
\end{center}

\renewcommand{\thefootnote}{\arabic{footnote}}

\begin{center}
{\it   \footnotesize
Department of Physics, University of North Carolina,\\
Chapel Hill, NC 27599, USA}
\end {center}

\noindent
\begin{center} {\bf Abstract} \end{center}
We consider the Type IIB string theory in the presence of various extra
$7/\overline 7$-brane pairs compactified on a warped Calabi-Yau threefold 
that admits a conifold singularity. We demonstrate that the volume modulus 
can be stabilized perturbatively at a non-supersymmetric $AdS_4/dS_4$ vacuum 
by the effective potential that includes the stringy $(\alpha^\prime)^3$ 
correction obtained by Becker {\it et al.} together with a combination 
of positive tension and anomalous negative tension terms generated by the 
additional 7-brane-antibrane pairs.
\vfill

\begin{flushright}
   December, 2004
\end{flushright}
\end{titlepage}

\bigskip

\hfill{}
\bigskip

\section{Introduction}
\label{sec:intro}
Moduli stabilization in sting theory using warped compactifications \cite{Chan:2000ms}
with nontrivial fluxes has proven to be of great success
\cite{Giddings:2001yu}, \cite{Kachru:2002he}, \cite{Berg:2003np}, \cite{Berg:2003ri}, \cite{Tripathy:2002qw}, \cite{Frey:2002hf}, \cite{Ashok:2003gk}, \cite{Denef:2004ze}, \cite{Giryavets:2004zr}, \cite{DeWolfe:2002nn}, \cite{Kachru:2004jr}.
In the case of Type IIB warped compactifications on an orientifolded Calabi-Yau with background 
NS and RR three-form fluxes, all complex-structure moduli as well as the dilaton can be stabilized.
Because of the ``no-scale'' structure of the resulting supergravity potential, the K\"ahler moduli 
remain unfixed. It was argued that $\alpha^\prime$ stringy corrections and non-perturbative effects 
may break the ``no-scale'' structure providing a mechanism to freeze the K\"ahler moduli \cite{Giddings:2001yu}. 
It has subsequently been demonstrated that the non-perturbative contributions to the 
superpotential  \cite{Witten:1996bn}, \cite{Katz:1996th} can indeed stabilize
the volume modulus producing a supersymmetric $AdS_4$ vacuum \cite{Kachru:2003aw}. 
Moreover, by adding extra sources such as anti-D3-branes one can brake
supersymmetry and tune the fluxes to lift the $AdS_4$ ground state to obtain a 
metastable $dS_4$ \cite{Kachru:2003aw}. Getting de Sitter space from string theory
has been a rather difficult problem. In particular, the ``no-go'' theorem 
\cite{Maldacena:2000mw}, \cite{deWit:1986xg} 
states that a de Sitter solution cannot be obtained in string or M-theory if one 
only uses the low-energy supergravity action. Sigma model $\alpha^\prime$
corrections, perturbative effects in $g_s$ and inclusion of extended sources were expected to
violate the ``no-go'' conditions which was successfully demonstrated by an explicit
construction \cite{Kachru:2003aw}. 
A different way to obtain a $dS$ space solution in string theory was found earlier in 
\cite{Silverstein:2001xn}, \cite{Maloney:2002rr} where some non-geometric
effects at the string scale were used. The scenario of Kachru, Kallosh, Linde and 
Trivedi (KKLT) \cite{Kachru:2003aw} has been more succesfull since
it incorporates many attractive model-building features such as a branes moving
in a warped geometry which may provide a possible solution to the hierarchy problem
in a context of a brane world model. Yet, more generic models can probably be constructed
at the string scale so it becomes a matter of time to see which scenario prevails in the end. 
Variations of the original KKLT model have also been found \cite{Burgess:2003ic}, \cite{Saltman:2004sn}. 
For constructions of de Sitter solutions in Heterotic string theory see \cite{Becker:2004gw},
\cite{Buchbinder:2004im}. In a recent work, the authors of
\cite{Balasubramanian:2004uy} suggested to modify the KKLT proposal by incorporating stringy
corrections to the K\"ahler potential \cite{Becker:2002nn} to obtain non-supersymmetric
vacua including $dS_4$. A new alternative to the KKLT construction of de Sitter vacua appeared
recenlty in \cite{Saltman:2004jh} where flux compactifications on products of Riemann 
surfaces are considered and the leading effects stabilizing the moduli are perturbative.
For good reviews see \cite{Frey:2003tf}, \cite{Silverstein:2004id}, \cite{Balasubramanian:2004wx}.
Many inflationary models motivated by the KKLT scenario have been proposed \cite{Kachru:2003sx}, 
\cite{Blanco-Pillado:2004ns}, \cite{Denef:2004dm}, \cite{Dasgupta:2004dw}, \cite{Burgess:2004kv},
\cite{Kallosh:2003ux}, \cite{Kallosh:2004yh}, \cite{Firouzjahi:2003zy}, \cite{Hsu:2003cy},
\cite{Hsu:2004hi}, \cite{Buchel:2004qg}, \cite{Buchel:2003qj}, \cite{Kallosh:2004yh},
\cite{Shandera:2004aa}.
In a recent paper \cite{Kallosh:2004yh} the authors discuss some generic problems
that exist in the simplest version of the KKLT motivated inflationary models.
In particular, they point out at the restriction that relates the inflationary
Hubble scale and SUSY breaking scale $H \lesssim m_{3/2}$ which implies that in a
simplest version of KKLT based inflation, high energy scale of SUSY braking is favored.
They then propose a more general setup based on racetrack type superpotentials where 
this restriction can be avoided altogether.
Apart from the above generic problem, a different type of obstacle appears 
in the brane-antibrane inflation realized in warped geometry \cite{Kachru:2003sx}. 
This problem is related to the superpotential
volume stabilization mechanism used in the KKLT scenario. More specifically, the
authors of \cite{Kachru:2003sx} argue that for a generic functional
form of the superpotential, the inflaton potential gets modified in a way that makes it too
steep for inflation. One possible resolution to this problem briefly discussed
in \cite{Kachru:2003sx} would be K\"ahler volume stabilization mechanism. This method would
allow one to stabilize the volume modulus directly without generating a large inflaton mass.
On the other hand, in the case of superpotential volume stabilization volume modulus
is a component of a superfield which is stabilized. This is the key difference that
makes K\"ahler volume stabilization compatible with the brane-antibrane inflation.

We consider a Type IIB theory warped compactification with non-trivial NS 
and RR fluxes on a Calabi-Yau threefold ${\cal M}$ that admits a conifold 
singularity with various 7-branes wrapped on the four-cycles inside ${\cal M}$.
Supersymmetry is broken by a non-zero $(0,3)$ component of the $G_{(3)}$ 
flux that generates a Gukov-Vafa-Witten \cite{Gukov:1999ya} superpotential $W\neq 0$. 
In the F-theory picture this is a compactification on an elliptically-fibered 
Calabi-Yau fourfold $X$ with a Calabi-Yau threefold ${\cal M}$ as the base of 
the fibration and the 7-branes embedded into the base at the special loci 
where the fiber $T^{\,2}$ degenerates. All complex-structure moduli
are stabilized by a choice of the integer fluxes \cite{Giddings:2001yu}.

We show that the volume modulus as well as the dilaton can be stabilized 
perturbatively when we combine the stringy corrections to the ``no-scale'' 
potential obtained in \cite{Becker:2002nn} with the contributions induced by 
additional pairs of (p,q) $7/\overline7$-branes and ${\rm D7/\overline {D7}}$-branes
wrapped on the four-cycles\footnote{Our construction is a hybrid
that features a warped compactification geometry as in \cite{Kachru:2003aw} but
uses a perturbative mechanism for volume stabilization similar to 
\cite{Saltman:2004jh}.}.
Apart from the positive tension terms which dominate at large volume,
the extra brane-antibrane pairs also generate an anomalous 
negative tension term proportional to the triple intersection numbers 
that has a subdominant large volume dependence. We also assume that the loci where 
the $T^{\,2}$ fiber degenerates are located far from the highly warped region so 
the warp factor scaling the potential energy due to the $7$-branes is trivial 
and the volume dependence is not modified by the warping.
When the Euler characteristic of the Calabi-Yau is negative $\chi({\cal M}) < 0$ 
the perturbative stringy correction yields a positive potential that dominates at small volume
over the other contributions in the effective potential.

By discretely varying the integer fluxes, the Euler characteristic 
and the number of extra brane-antibrane pairs in the effective 
potential we obtain a ``landscape'' of metastable de Sitter, non-supersymmetric 
Minkowski and anti-de Sitter vacua where supersymmetry is broken at the Kaluza-Klein 
scale. Such flexibility in the choice of flux, topological and brane 
numbers implies that we can stabilize the volume at a large value and the string 
coupling $g_s$ at a small value, providing theoretical control over the 
perturbative expansion in the low-energy effective field theory. Such control is necessary 
in the absence of low-energy supersymmetry as loop corrections at various orders 
modifying the moduli potential will appear \cite{Saltman:2004jh}. 
In addition, in this regime, the non-perturbative corrections to the superpotential are 
exponentially suppressed and can be safely neglected.
\section{Review of Type IIB warped compactifications with fluxes}
\label{sec:second}
In Type IIB compactifications on a Calabi-Yau threefold $\cal M$ with
nontrivial NS and RR fluxes $H_{(3)}$ and $F_{(3)}$ \cite{Giddings:2001yu},
\cite{Kachru:2002he}, \cite{Grimm:2004uq} 
the fluxes induce warping of the background giving a warped product of Minkowski
space and the Calabi-Yau:
\be
\label{metric}
ds^2=e^{2A(y)}\eta_{\mu\nu}dx^\mu dx^\nu + e^{-2A(y)}g_{mn}dy^mdy^n,
\ee
where $g_{mn}$ is the Calabi-Yau metric. To construct such solution we
are forced to include extended local sources to cancel the D3-brane charge
produced by the fluxes following from the Bianci identity
\be
\label{bianci}
d\tilde F_{(5)}=H_{(3)}\wedge F_{(3)}.
\ee
In the language of F-theory compactified on an elliptically fibered Calabi-Yau 
fourfold {\it X}, to satisfy flux conservation (\ref{bianci}) we can introduce
various 7-branes sitting at the special loci in the base 
$\cal M$ where the elliptic fiber $T^{\,2}$ degenerates.
The 7-branes wrap the four-cycles in $\cal M$ generating an anomalous 
negative D3-brane charge induced by the curvature of the four-cycles
$Q_3^{7}=- \chi ({\it X})/24 $, where $\chi ({\it X})$ is the 
Euler characteristic of the fourfold. In this language, the varying profile 
of the Type IIB axion-dilaton $\tau\equiv C_{(0)}+ie^{-\phi}$ corresponds 
to the complex structure of the elliptic fiber.
In the limit of the Type IIB Calabi-Yau orientifold, we can introduce a set
of $N_{\rm O3}$ orientifold O3-planes which quotient the space by a discrete 
symmetry and carry a D3-brane charge $Q_3^{\rm O3}=- N_{\rm O3}/4$.
The total D3-brane charge required to vanish is then
\be
\label{charge_cons}
Q_3^{local}+\frac {1}{2\kappa^2_{10}T_3}
\int_{\cal M} H_{(3)}\wedge F_{(3)}=0,
\ee
where $Q_3^{local}$ includes contributions from all local sources of the D3-brane charge:
7-branes or O3-planes, mobile D3 and $\overline {\rm D3}$-branes.
The three-form fluxes induce a Gukov-Vafa-Witten (GVW) superpotential 
\cite{Gukov:1999ya} for the compex-structure moduli
\be
\label{gvw}
W=\int_{\cal M}\Omega\wedge G_{(3)},
\ee
where $G_{(3)}\equiv F_{(3)}-\tau H_{(3)}$ and $\Omega$ is the holomorphic three-form
of the Calabi-Yau.
The Calabi-Yau threefold $\cal M$ in general has a large number of both complex-structure
moduli $z_a$ and K\"ahler moduli $\rho_i$ as well as the axion-dilaton $\tau$.
The geometric moduli surviving the orientifold projection include $h^{1,1}_+$ even
K\"ahler deformations and $h^{1,2}_-$ odd complex-structure deformations\footnote{See a recent work \cite{Grimm:2004uq} on constructing the effective action for {\cal N}=1 Type IIB
on Calabi-Yau orientifolds.}. The axion-dilaton also survives the projection.
In further analysis we will assume a large volume limit and
later check it for consistency. Here we will use units where the four-dimensional
Newton's constant $\kappa^2_4=(8\pi G_N)=1$.
For the simple case of a single K\"ahler modulus $\rho$, where
${\rm Im}\rho=\hat\sigma=e^{4u-\phi}$ corresponds to the fluctuations of the overall
volume ${\cal V}=e^{6u}$, the tree-level K\"ahler potential takes the following form:
\be
\label{kahler}
K=-3{\rm log}\left(-i(\rho-\bar\rho)\right)-{\rm log}\left(-i
\int_{\cal M}\Omega\wedge\overline\Omega\right)
-{\rm log}\left(-i(\tau-\bar\tau)\right).
\ee
A more general case with several K\"ahler moduli and a nontrivial
warp factor is considered in \cite{DeWolfe:2002nn}.
The standard ${\cal N}=1$ supergravity potential is given by
\be
\label{pot}
V=e^K\left({\cal G}^{A{\bar B}}D_AW\overline{D_{\bar B}W}-3|W|^2\right),
\ee
where $D_AW=\partial_AW + W\partial_A K$ and ${\cal G}_{A\bar B}=\partial_A\partial_{\bar B} K$.
The structure of the K\"ahler potential (\ref {kahler}) is such that
when we use it to compute the potential  (\ref {pot}) the $-3|W|^2$ term cancels
out and the potential reduces to the familiar ``no-scale'' form \cite{Giddings:2001yu}
\be
\label{noscalepot}
V=e^K\left({\cal G}^{a\bar b}D_aW\overline{D_{\bar b}W}+
{\cal G}^{\tau\bar\tau}D_\tau W\overline{D_{\bar\tau}W}\right).
\ee
Notice, that $A,\bar B$ in (\ref {pot}) run over all moduli, whereas
in (\ref{noscalepot}) the indices run only over the complex-structure moduli
$z_a,\bar z_b$ and axion-dilaton $\tau$. This implies that even after we
freeze $z_a{\rm 's}$ and $\tau$, the K\"ahler modulus $\rho$ remains unfixed.
However, see \cite{Saltman:2004sn} for an interesting way around this problem
where nonzero local minima of the no-scale potential were found.
One of the conditions to obtain a warped solution (\ref{metric}) is the
requirement that the three-form $G_{(3)}$ must be imaginary self-dual\cite{Giddings:2001yu}:
\be
\label{isd}
*_6G_{(3)}=iG_{(3)},
\ee
which is equivalent to the following requirements
\bea
\label{eqv}
&&D_\tau W=0,\\
&&D_a W=0.\nonumber
\eea
The Klebanov-Strassler (KS) solution \cite{Klebanov:2000hb} is an example 
of such warped
geometry where the six-dimensional internal space is a warped deformed
conifold with a tip smoothed by an $S^3$.
The fluxes satisfy the following quantization conditions
\be
\label{quantcond}
\frac{1}{(2\pi)^2\alpha^\prime}\int_AF_{(3)}=M,\,\,\,\,\frac{1}{(2\pi)^2\alpha^\prime}
\int_BH_{(3)}=-K,
\ee
where A is a shrinking $S^3$ at the tip of the conifold and B is a dual cycle
giving the following GVW superpotential \cite{Giddings:2001yu}:
\be
\label{sup}
W=(2\pi)^2\alpha^\prime\left[M{\cal G}(z)-K\tau z\right],
\ee
where $z$ is a complex coordinate on the collapsing cycle defined by
\be
\label{zdef}
z=\int_A\Omega,
\ee
and on the dual cycle
\be
\label{dualdef}
\int_B\Omega\equiv{\cal G}(z)=\frac{z}{2\pi i}{\rm ln}z+{\rm holomorphic}.
\ee
Once the integer values $M$ and $K$ in (\ref{quantcond}) are fixed and
condition (\ref{isd}) is satisfied, all complex structure moduli $z_a$
are frozen. The dilaton can also be fixed if we turn on fluxes on another
pair of three cycles $A^\prime,B^\prime$.
The warp factor has a minimum value at the bottom of the throat given by
\be
\label{wfactor}
e^{A_{\rm min}}\sim z^{1/3}\sim {\rm exp}\left(-2\pi K/3Mg_s\right),
\ee
creating a hierarchy of energy scales determined by the choice of the integer
fluxes.
In a generic case supersymmetry is broken by a non-vanishing $(0,3)$ part
of $G_{(3)}$ that induces a constant non-zero value for
the superpotential $W=W_0$. Requiring unbroken supersymmetry
amounts to requiring that $D_{\rho}W=0$, which implies that
the superpotential $W=0$ since it is independent of $\rho$.
The authors of \cite{Kachru:2003aw} proposed a mechanism to stabilize the K\"ahler
modulus by including $\rho$-dependent non-perturbative corrections to the
tree-level superpotential \cite{Witten:1996bn}, \cite{Katz:1996th}, \cite{Berg:2004ek}, \cite{Berg:2004sj}:
\be
\label{nonpert}
W=W_0+Ae^{ia\rho}.
\ee
In this way the ``no-scale'' structure of (\ref{noscalepot}) gets broken
resulting in a potential with a supersymmetric $AdS_4$ minimum where
the volume modulus is stabilized at a large value. A further step was then taken
to introduce anti-D3-branes that brake supersymmetry and generate, due to their tension,
a positive potential which dominates at large volume.
By varying the number of anti-D3-branes and integer fluxes, the minimum in the effective 
potential can be lifted to obtain a discrete ``landscape'' of metastable de 
Sitter vacua \cite{Kachru:2003aw}.
\section{Volume stabilization and de Sitter space}
\label{sec:third}
\subsection{BBHL stringy corrections}
\label{subsec:one}
The authors of \cite{Giddings:2001yu} suggested that the ``no-scale'' structure
of the potential (\ref{noscalepot}) may be broken in the quantum theory by
higher order stringy corrections. Such higher order $(\alpha^\prime)^3$
corrections to the potential were explicitly computed by
Becker, Becker, Haack, and Louis (BBHL) \cite{Becker:2002nn} based on the previous
work \cite{Gross:1986iv}, \cite{Freeman:1986br}, \cite{Grisaru:1986kw} in the context
of a Type IIB warped compactification with three-form fluxes.
The new terms found in \cite{Becker:2002nn} originate from the correction of the 
K\"ahler
moduli part of the K\"ahler potential which in string frame reads:
\bea
\label{modkahler}
K&=&-2{\rm log}\left[2{\cal V} e^{-3\phi_0/2}+\xi\left(\frac{-i(\tau-\bar\tau)}{2}
\right)^{3/2}\right]
-{\rm log}\left(-i\int_{\cal M}\Omega\wedge\overline\Omega\right)\\
&-&{\rm log}\left(-i(\tau-\bar\tau)\right),\nonumber
\eea
while the flux induced superpotential (\ref{gvw}) remains uncorrected
\cite{Giddings:2001yu}, \cite{Witten:1985bz}.
The term in the first line in (\ref{modkahler}) contains
the dilaton-dependent $(\alpha^\prime)^3$ correction proportional to
\be
\label{xi}
\xi=-\frac{\chi({\cal M})}{2}\zeta(3),
\ee
where $\chi({\cal M})$ is the Euler characteristic of the Calabi-Yau threefold
and we used the convention of \cite{Becker:2002nn} to set $2\pi\alpha^\prime=1$.
In their computation, the authors of \cite{Becker:2002nn} did not include the warp
factor $e^{4A}$. It would be interesting to see if including it would change
our further conclusions. For now we will assume that we consider a large
volume limit so the warp factor can be ignored. The volume modulus $\cal V$
can be expressed as an implicit function of the volumes of the four-cycles
$\sigma_i$ in terms of the areas of the two-cycles $v^k$ and in the large
volume limit is given by:
\bea
\label{volumenot}
&&{\cal V}=\int_{\cal M}J^3=\frac {1}{6}k_{ij\,k}v^iv^jv^k,\\
&&\sigma_i=\frac {1}{6}k_{ij\,k}v^jv^k,\nonumber
\eea
where $J$ is the K\"ahler form and $k_{ij\,k}$ are constant
intersection numbers.\footnote{The indices are raised with $\delta^{ij}$.}
The physical volume of the Calabi-Yau
\be
\label{vol}
{\rm V}_{CY}\equiv{{\cal V}}(2\pi\alpha^\prime)^3={\cal V}
\ee
In the four-dimensional Einstein frame the Kahler moduli are rescaled as
\bea
\label{vi}
&&\hat v_i=v_ie^{-\phi_0/2},\\
&&\hat\sigma_i=\sigma_ie^{-\phi_0},\nonumber
\eea
so $\hat{\cal V}={\cal V}e^{-{3\phi_0}/{2}}$.
In order to keep using the same notation as in Section \ref{sec:second}, we will identify
the complexified K\"ahler moduli as $\rho_i=\frac{1}{3}g_i+i\hat\sigma_i$, where
$g_i$ is the axion coming from the KK-reduction of the RR four-form.
Including the $(\alpha^\prime)^3$ corrections in the K\"ahler potential breaks the
``no-scale'' structure resulting in the following form of the potential \cite{Becker:2002nn}:
\bea
\label{modpot}
V&=&{e^K}\left({\cal G}^{a\bar b}D_aW\overline{D_{\bar b}W}+
{\cal G}^{\tau\bar\tau}D_\tau W\overline{D_{\bar\tau}W}\right)
-9\frac{\hat\xi\hat{\cal V}e^{-\phi_0}}{(\hat\xi -\hat{\cal V})(\hat\xi +2\hat{\cal V})}\\
&\times&{e^K}\left(W\overline{D_{\bar\tau}W}+\overline WD_\tau W\right)
-3\hat\xi\,\frac {(\hat\xi^2+7\hat\xi\hat{\cal V}+\hat{\cal V}^2)}
{(\hat\xi -\hat{\cal V})(\hat\xi +2\hat{\cal V})^2}e^K|W|^2,\nonumber
\eea
where $\hat\xi=\xi e^{-3\phi_0/2}$ and the components of the inverse metric
necessary to compute (\ref{modpot}) are:
\bea
\label{metcomp}
&&{\cal G}^{\tau\bar\tau}=e^{-2\phi_0}\frac{4\hat{\cal V}-\hat\xi}{\hat{\cal V}-\hat\xi},\\
&&{\cal G}^{\tau\bar\rho_i}=e^{-\phi_0}\frac{3\hat\xi}{\hat{\cal V}-\hat\xi}\hat\sigma_i,\nonumber\\
&&{\cal G}^{\rho_i\bar\rho_j}=-\frac{2}{9}(2\hat{\cal V}+\hat\xi)k_{ij\,k}\hat v^k+
\frac{4\hat{\cal V}-\hat\xi}{\hat{\cal V}-\hat\xi}\hat\sigma_i\hat\sigma_j.\nonumber
\eea
Once we fix the integer fluxes and impose conditions (\ref{eqv}) necessary to
obtain a warped solution (\ref{metric}), all complex-structure moduli $z_a$
become fixed and potential (\ref{modpot}) takes the following form:
\be
\label{finalpot}
V= 3{\hat\xi}\,\frac {(\hat\xi^2+7\hat\xi\hat{\cal V}+\hat{\cal V}^2)}{(\hat{\cal V}-\hat\xi)
(2\hat{\cal V}+\hat\xi)^2}e^{K}|W|^2,
\ee
Notice that the potential is proportional to the square of the flux induced superpotential
(\ref{gvw}) and therefore vanishes unless supersymmetry is broken
by the $(0,3)$ part of the three-form flux $G_{(3)}$.
\subsection{Volume modulus stabilization}
\label{subsec:two}
The authors of \cite{Becker:2002nn}
rightly point out that the result in (\ref{finalpot}), containing all orders in $\alpha^\prime$
beginning with $(\alpha^\prime)^3$, can only be trusted to order $(\alpha^\prime)^3$. In fact,
to this leading order this correction to the ``no-scale'' potential is exact. 
Thus, keeping the leading order terms, we obtain the following potential to order $(\alpha^\prime)^3$:
\be
\label{correction}
V_s=\frac{3\hat\xi}{4\hat{\cal V}}e^{K^{(0)}}|W|^{\,2}
=-A\frac{\chi|W|^{\,2}}{2}\left(\frac{g_s^4}{{\cal V}^{\,3}}\right),
\ee
where $K^{(0)}$ is the supergravity K\"ahler potential
\be
\label{kahlernot}
K^{(0)}=\phi_0-2{\rm log}(\hat{\cal V})
-{\rm log}\left(-i\int_{\cal M}\Omega\wedge\overline{\Omega}\right)+{\rm const.},
\ee
and $A$ is a numerical constant. In the four-dimensional Einstein frame
this type of classical higher derivative correction should scale as 
$g_s^2$ in string coupling. Normally, this is indeed the case, 
since the superpotential $W$ is linear in $\tau$ and ${\rm Im}(\tau)=1/g_s$
so the factor of $|W|^2$ in (\ref{correction}) gives the expected scaling.
However, in the case of a highly warped geometry at the bottom of the
Klebanov-Strassler throat, the GVW superpotential (\ref{sup})
is effectively given by
\be
\label{redsup}
W\sim(2\pi)^2\alpha^\prime M.
\ee
Here we used the fact that the holomorphic part of ${\cal G}(0)\sim O(1)$
and the second term in (\ref{sup}) containing $\tau$ is multiplied by 
$z$ and therefore is exponentially suppressed as we take $K/(g_s M)\gg 1$
to obtain a large hierarchy of scales in the highly warped region.
Unfortunately, this type of potential by itself still cannot fix the
volume modulus because of its runaway behaviour at large volume
where it vanishes\footnote{The authors of \cite{Balasubramanian:2004uy} combined 
the perturbative stringy $(\alpha^\prime)^3$ correction with non-perturbative contributions
to the superpotential (\ref{nonpert}) to find non-supersymmetric $AdS_4/dS_4$ vacua.}.
However, recall that in our compactification scheme we had
to include certain extended sources of the D3-brane charge in order
to cancel the tadpole anomaly (\ref{charge_cons}). In the F-theory language
such objects were various 7-branes wrapped on the four-cycles inside the base ${\cal M}$, 
that induced an anomalous D3-brane charge $-\chi(X)/24$, where $\chi(X)$ is
the Euler characteristic of the Calabi-Yau fourfold $X$. In case of
low-energy supersymmetry, this anomalous negative contribution to the D3 charge 
is related to the negative tension by a BPS condition. The corresponding 
negative energy contribution is then cancelled by one of the positive
terms due to the fluxes making the total energy positive semi-definite \cite{Giddings:2001yu}. 
We will therefore use the same strategy that was used by Silverstein\footnote{I would like to 
thank Eva Silverstein for explaining some key points related to this part of the paper.} 
and Saltman in the recent work \cite{Saltman:2004jh}. We will introduce $n_{D7}$ 
additional pairs of $D7/\overline {D7}$-branes and $n_7$ extra pairs of $(p,q)$ 
$7/\overline 7$-branes wrapped on the four-cycles in ${\cal M}$ placed at the 
loci where the fiber $T^2$ degenerates.
These pairs induce no net D3-brane charge so the Gauss law (\ref{charge_cons}) is still 
satisfied. In fact, in a non-supersymmetric case considered here, a configuration 
that includes additional brane-antibrane pairs is more generic\footnote{In \cite{Saltman:2004jh} 
the authors used $n_7$ additional sets of 24 (p,q) 7-branes whose contributions are
known explicitly from F-theory on K3.}. These extra local sources 
generate an anomalous negative D3-brane tension which can be tuned to a 
large value independently of the total D3-brane charge which gives us a lot 
of extra flexibility. In particular, in Einstein frame, these extra brane-antibrane
pairs make the following total contribution to the effective potential \cite{Saltman:2004jh}:
\be
\label{o3pot}
V_7\sim -N_7\left(\frac{g_s^3}{{\cal V}^{\,2}}\right)+n_7\left(\frac{g_s^2}{{\cal V}^{4/3}}\right)
+n_{D7}\left(\frac{g_s^3}{{\cal V}^{4/3}}\right),
\ee
where $N_7$ is an effective parameter given in terms of triple intersections of 
7-branes and is proportional to qubic combinations of $n_7$ and $n_{D7}$. 
Here we have assumed that the points in the base ${\cal M}$ where the F-theory
elliptic fiber degenerates are located far from the highly warped region.
Therefore we have set the warp factor scaling the 7-brane tensions to unity $e^{4A}\approx 1$. 
Notice that the negative contribution is qubic in $n_7$ and $n_{D7}$ and
dominates at small volume over the other two terms. This large negative energy 
makes annihilation of brane-antibrane pairs energetically disfavored.
One important step needs to be done before we combine contributions (\ref{correction})
and (\ref{o3pot}). Recall that in (\ref{correction}) the volume is measured in units
of $(2\pi\alpha^\prime)^3$ as in (\ref{vol}) whereas in (\ref{o3pot}), the volumes
are given in units of $(\alpha^\prime)^3$ which we will use in the total 
effective potential. Thus, going to the latter unit convention we obtain the
following effective potential up to factors of order one
\be
\label{total}
V_t= - \chi(2\pi)^{13}M^{\,2}\left(\frac{g_s^4}{{\cal V}^{\,3}}\right)
-N_7\left(\frac{g_s^3}{{\cal V}^{\,2}}\right)+n_7\left(\frac{g_s^2}{{\cal V}^{4/3}}\right)
+n_{D7}\left(\frac{g_s^3}{{\cal V}^{4/3}}\right),
\ee
where we also used (\ref{redsup}) to substitute for $W$ and set $\alpha^\prime=1$. Notice 
that if the Calabi-Yau manifold has a negative Euler characteristic $\chi ({\cal M})<0$, 
which is the case if the number of complex-structure deformations is larger than the number 
of K\"ahler deformations, the first term in (\ref{total}) becomes positive. It dominates
over the other terms at small volume and therefore the total effective potential features a minumum
at a finite volume. By discretely varying the fluxes, the Euler characteristic and numbers of 
brane-antibrane pairs we obtain a ``landscape'' of metastable de Sitter, non-supersymmetric
Minkowski and anti-de Sitter vacua. To find the minimum of (\ref{total}) we can first take a
derivative with respect to the string coupling and by setting the derivative to zero, 
we can find $g_s$ in terms of ${\cal V}$. In fact, this is relatively easy since in the end
it boils down to solving a simple quadratic equation that gives the following
\be
\label{solution}
g_s=-\frac{3 N_7{\cal V}-3n_{D7}{\cal V}^{5/3}+
\Delta^{1/2}}{8\chi(2\pi)^{13}M^{\,2}},
\ee
where the discriminant $\Delta$ is given by
\be
\label{discrim}
\Delta=32\chi(2\pi)^{13}M^{\,2}n_7{\cal V}^{5/3}+9\left(n_{D7}
{\cal V}^{5/3}-N_7{\cal V}\right)^2.
\ee

\begin{figure}[ht]
\begin{center}
\leavevmode
\epsfxsize 11 cm
\epsfbox{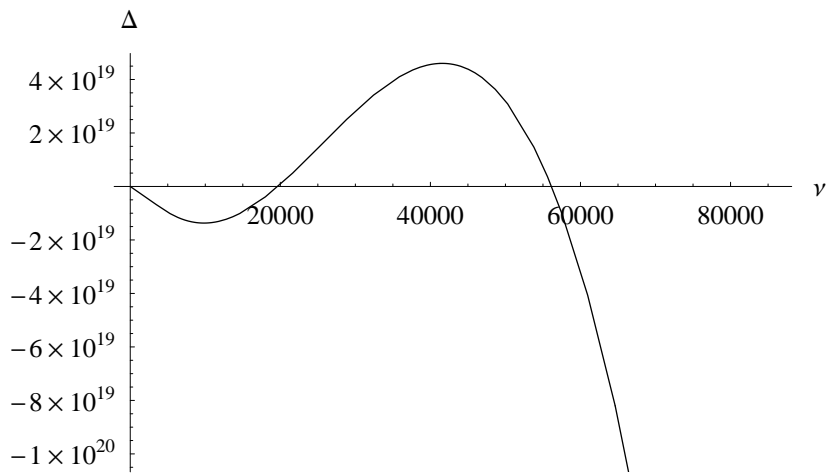}
\end{center}
\caption{ \footnotesize
Discriminant (\ref{discrim}) plotted as a function of volume for the set of parameter values 
chosen above. This plot demonstrates the excluded volume range where $\Delta <0$.
The discriminant eventually again turns positive as we go to extremely large values of 
${\cal V}$ which are not included in this plot.}
\label{discriminantplot}
\end{figure}

\begin{figure}[ht]
\begin{center}
\leavevmode
\epsfxsize 10 cm
\epsfbox{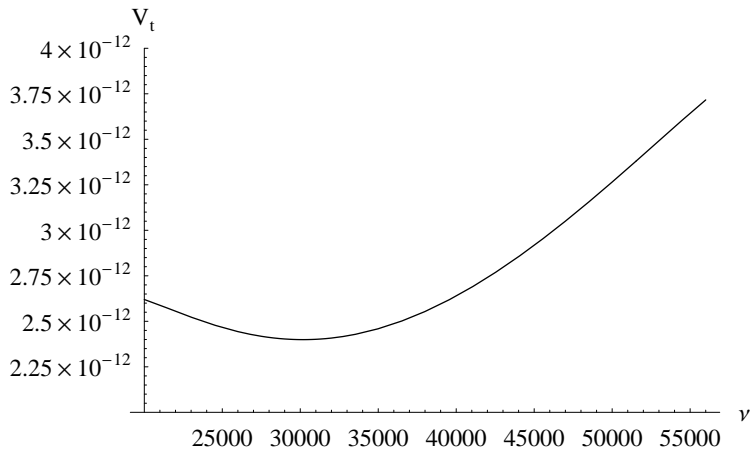}
\end{center}
\caption{ \footnotesize
Total effective potential plotted for the volume range where the discriminant $\Delta>0$.
For our particular choice of flux and brane numbers, the volume is
stabilized at a large value ${\cal V}_{min}\sim 3\times 10^4(\alpha^\prime)^3$ and
the minimum is de Sitter, given by $V_{dS}\sim 2.4\times 10^{-12}$.}
\label{potentialplot}
\end{figure}

Since we are considering the case when $\chi<0$, the above solution 
can yield an imaginary part when the discriminant turns negative.
This happens for small volumes since the first term in (\ref{discrim}) always
dominates at the start. It also takes place in the range of volumes where 
the terms in the brackets are nearly cancelled. Keeping these restrictions
in mind, we can plug (\ref{solution}) into (\ref{total}) to obtain the
potential as a function of volume. In order to demonstrate how the available 
parameters can be tuned to give a de Sitter vacuum located at large volume 
with the string coupling stabilized at a small value we construct a simplified 
toy model where we set $N_7=n_7^3+n_{D7}^3$ and choose $\chi=-4$, $M=3$, $n_7=1$ 
and $n_{D7}=73$. The discriminant plotted in Fig. \ref{discriminantplot}
as a function of volume yields the volume range where the solution for the
string coupling (\ref{solution}) is real. 
Thus, we can plug the string coupling 
(\ref{solution}) into the total effective potential (\ref{total}) and plot 
it as a function of volume in the allowed range Fig.\ref{potentialplot}.

\begin{figure}[ht]
\begin{center}
\leavevmode
\epsfxsize 9.7 cm
\epsffile{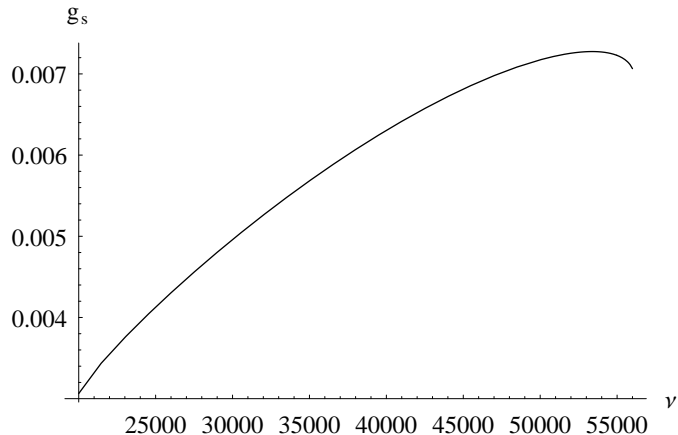}
\end{center}
\caption{ \footnotesize
String coupling as a function of stabilized volume plotted in the range where $\Delta >0$.}
\label{couplingplot}
\end{figure}

\begin{figure}[ht]
\begin{center}
\leavevmode
\epsfxsize 15 cm
\epsffile{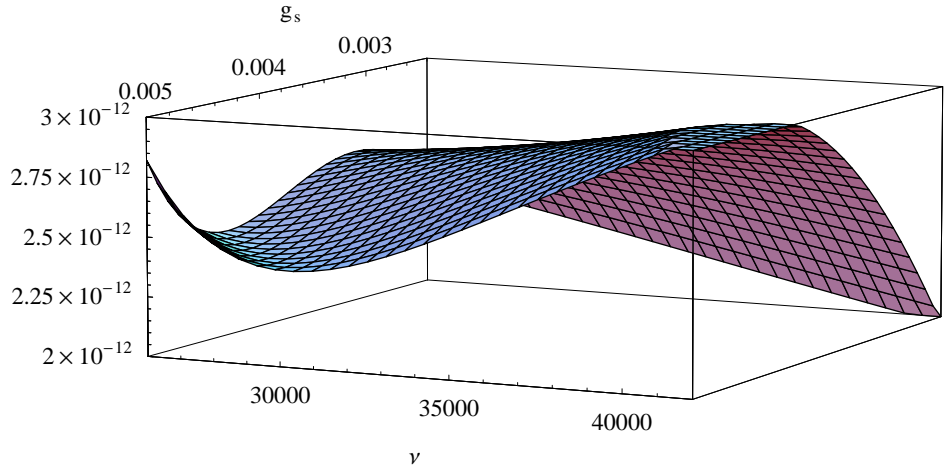}
\end{center}
\caption{ \footnotesize
Effective potential (\ref{total}) plotted as a function of string coupling and volume. 
The values that the volume and the string coupling can take are not restricted.
The metastable de Sitter vacuum can be tuned to a desired value by discretely varying 
flux, topological and brane numbers.}
\label{3dplot}
\end{figure}

We find that the volume is stabilized at a large value ${\cal V}_{min}\sim 3\times 10^4(\alpha^\prime)^3$
where the potential has a de Sitter minimum $V_{dS}\sim 2.4\times 10^{-12}$. In order to
find the stabilized value for $g_s$ we plug ${\cal V}_{min}$ into (\ref{solution})
and find that the string coupling is fixed at a value $g_s\sim 5\times 10^{-3}$.
The string coupling in (\ref{solution}) plotted as a function of volume for the same 
range where $\Delta>0$ is given in Fig.\ref{couplingplot}.
This apparent restriction on the allowed values of ${\cal V}$ in the plots above is an
artifact created by the method which we used to solve for the minimum. Of course, when we
plot the effective potential (\ref{total}) in 3D as a function of ${\cal V}$ and $g_s$, they
can vary over all positive values Fig.\ref{3dplot}.
The values ${\cal V}_{min}$ and $g_s$ obtained for our toy model satisfy both
consistency checks. First, we demonstrated that the volume modulus can be
stabilized at a large value and our reliance on the BBHL result that
does not take into account the warp factor is valid. Second, we showed
that the string coupling can be stabilized at a small enough value.
Recall that the non-perturbative corrections to the tree-level superpotential
of the type (\ref{nonpert}) are highly suppressed in the weak coupling limit
and therefore we can safely ignore them in our construction. To make sense of the stabilized
volume dependence on the number of integer fluxes, recall that the three-cycle at the 
tip of the conifold whose size is given by the minimal warp factor in (\ref{wfactor}) 
tries to expand to minimize the energy density created by $M$ units of RR flux 
$F_{(3)}$. Thus, as we pump the energy into the three-cycle at the tip by 
increasing the integer $M$ the overall volume also grows.
Similarly to \cite{Saltman:2004jh}, supersymmetry is broken at the Kaluza-Klein scale.
Serving the role of supersymmetry breaking order parameter is the gravitino 
mass given by
\be
\label{susy}
m_{3/2}\sim e^{K_0/2}|W|\sim (2\pi)^5 M\left(\frac{g_s^2}{{\cal V}_{min}}\right)
\sim 2.4\times 10^{-5}M_p\sim 6\times 10^{13}{\rm GeV}.
\ee
At such high SUSY breaking scale the flexibility provided by the choice of
flux, topology and brane numbers gives us the necessary theoretical control over
the perturbative expansion in the low-energy effective field theory.
In other words, we can tune the volume to a large value to minimize energy
densities and stabilize the coupling at a very small value which gives us
control over various higher order loop corrections.
Our construction can also include mobile D3 and $\overline{D3}$-branes.
Due to their interaction with the background fluxes, the $\overline{D3}$-branes 
naturally move to the highly warped region at the bottom of the KS throat and  
the corresponding term in the effective potential would have the same 
scalings as the last term in (\ref{total}). In this type of a brane-world model, the hierarchy 
of scales is then generated by the warp factor \cite{Chan:2000ms},
\cite{Giddings:2001yu}, \cite{DeWolfe:2002nn} as opposed to the low-energy 
supersymmetry\footnote{See \cite{Arkani-Hamed:2004fb},\cite{Arkani-Hamed:2004yi}
for some models with high SUSY breaking scale.}.
 
Our mechanism of volume stabilization has an important 
application for brane-antibrane inflationary models. The authors of 
\cite{Kachru:2003sx} consider a particular brane-antibane inflationary 
scenario in a warped compactification geometry inspired by the KKLT model \cite{Kachru:2003aw}.
In this case the warping provides for a very flat inflaton
potential naturally suitable for inflation. However, as
the authors in \cite{Kachru:2003sx} point out, the superpotential volume stabilization
based on \cite{Kachru:2003aw} creates a new problem. The D3-brane moduli
modify the definition of the imaginary part of the K\"ahler modulus $\rho$ in the following way:
\be
\label{voldef}
2\hat{\cal V}^{2/3}=-i(\rho-\bar\rho)-k(\phi,\bar\phi).
\ee
Since $\rho$ is a superfield, volume stabilization through non-perturbative
corrections to the superpotential fixes $\rho=\rho_0$ but not
the volume modulus ${\cal V}$. Since the D3-brane moduli $\phi,\bar\phi$
play the role of the inflaton field it is important that the
inflaton mass stays extremely small to allow for inflation with
enough e-foldings. Since in this case the volume of the Calabi-Yau
is not fixed directly the terms in the effective potential, when expanded around
$\rho_0$, result in a mass term for the inflaton. This yields a
slow-roll parameter $\eta=2/3$ which is too large for inflation to continue.
This problem can be alleviated by including $\phi$ dependence
into the superpotential and tuning the new terms that appear
to cancel the mass term. This solution does not seem very satisfactory
as it involves fine tuning to one percent level.
On the other hand, as discussed in \cite{Kachru:2003sx}, direct
volume stabilization mechanism via stringy corrections
to the K\"ahler potential demonstrated in this section,
allows to overcome this major obstacle. In this way the
volume modulus is stabilized directly so there is no
large inflaton mass term generated by the modification of the
potential (\ref{voldef}) due to the D3-brane moduli and the
brane-antibrane inflationary model in \cite{Kachru:2003sx} can be
realized without the need of fine tuning.
\section{Conclusions}
In this work we demonstrated that volume stabilization in a Type IIB warped Calabi-Yau
compactification with non-trivial NS and RR fluxes can be achieved 
perturbatively via a combination of 
stringy $\alpha^\prime$ corrections to the K\"ahler potential derived in \cite{Becker:2002nn} and
the contributions from additional 7-brane-antibrane pairs.
The volume modulus as well as the string coupling are stabilized at the minimum of the
corresponding effective potential provided the Euler characteristic of the
Calabi-Yau threefold is negative.
By varying the integer parameters such as the flux numbers, the Euler characteristic 
and the number of brane-antibrane pairs we obtain a discretuum of non-supersymmetric 
$AdS_4$, Minkowski and metastable $dS_4$ vacua in the 4d low-energy effective field theory.
Similarly to \cite{Saltman:2004jh}, in our compactification scheme 
supersymmetry is broken by the fluxes at the Kaluza-Klein scale.
In a toy example, we demonstrated that by tuning the flux, topology
and brane numbers, the string coupling can be stabilized at a small enough value
thus providing a handle on the effects from higher order terms appearing 
in the loop expansion. Likewise, the stabilized volume can be made large enough 
to ensure that energy densities stay small. In this way, theoretical control over the 
perturbative expansion in the low-energy effective field theory, necessary in 
the absence of low-energy supersymmetry is achieved.
Together with \cite{Saltman:2004jh}, where moduli stabilization for
a compactification on a product of Riemann surfaces was achieved perturbatively,
our construction featuring a Type IIB warped compactification on a Calabi-Yau
presents another alternative to the KKLT scenario \cite{Kachru:2003aw}. 
Apart from being very simple, the volume stabilization mechanism suggested here 
completely eliminates the large inflaton mass problem in the brane-antibrane 
inflationary model in warped geometry \cite{Kachru:2003sx}, which makes it
even more attractive.
\section* {Acknowledgements}
I would like to express my gratitude to Eva Silverstein for taking her
time to explain to me some important key issues relevant for this work.
I would also like to thank to Archil Kobakhidze, Heather Bobkova,
Louise Dolan and Laura Mersini for stimulating discussions and
to Andrei Linde and Ronen Plesser for answering some questions
related to this work.


\begin{thebibliography}{23}

\bibitem{Chan:2000ms}
C.~S.~Chan, P.~L.~Paul and H.~Verlinde,
``A note on warped string compactification,''
Nucl.\ Phys.\ B {\bf 581}, 156 (2000)
[arXiv:hep-th/0003236].

\bibitem{Giddings:2001yu}
S.~B.~Giddings, S.~Kachru and J.~Polchinski,
``Hierarchies from fluxes in string compactifications,''
Phys.\ Rev.\ D {\bf 66}, 106006 (2002)
[arXiv:hep-th/0105097].

\bibitem{Kachru:2002he}
S.~Kachru, M.~B.~Schulz and S.~Trivedi,
``Moduli stabilization from fluxes in a simple IIB orientifold,''
JHEP {\bf 0310}, 007 (2003)
[arXiv:hep-th/0201028].

\bibitem{Berg:2003np}
M.~Berg, M.~Haack and B.~Kors,
``Brane / flux interactions in orientifolds,''
Fortsch.\ Phys.\  {\bf 52}, 583 (2004)
[arXiv:hep-th/0312172].

\bibitem{Berg:2003ri}
M.~Berg, M.~Haack and B.~Kors,
``An orientifold with fluxes and branes via T-duality,''
Nucl.\ Phys.\ B {\bf 669}, 3 (2003)
[arXiv:hep-th/0305183].

\bibitem{Tripathy:2002qw}
P.~K.~Tripathy and S.~P.~Trivedi,
``Compactification with flux on K3 and tori,''
JHEP {\bf 0303}, 028 (2003)
[arXiv:hep-th/0301139].

\bibitem{Frey:2002hf}
A.~R.~Frey and J.~Polchinski,
``N = 3 warped compactifications,''
Phys.\ Rev.\ D {\bf 65}, 126009 (2002)
[arXiv:hep-th/0201029].

\bibitem{Ashok:2003gk}
S.~Ashok and M.~R.~Douglas,
``Counting flux vacua,''
JHEP {\bf 0401}, 060 (2004)
[arXiv:hep-th/0307049].

\bibitem{Denef:2004ze}
F.~Denef and M.~R.~Douglas,
``Distributions of flux vacua,''
JHEP {\bf 0405}, 072 (2004)
[arXiv:hep-th/0404116].

\bibitem{Giryavets:2004zr}
A.~Giryavets, S.~Kachru and P.~K.~Tripathy,
``On the taxonomy of flux vacua,''
JHEP {\bf 0408}, 002 (2004)
[arXiv:hep-th/0404243].

\bibitem{DeWolfe:2002nn}
O.~DeWolfe and S.~B.~Giddings,
``Scales and hierarchies in warped compactifications and brane worlds,''
Phys.\ Rev.\ D {\bf 67}, 066008 (2003)
[arXiv:hep-th/0208123].

\bibitem{Kachru:2004jr}
S.~Kachru and A.~K.~Kashani-Poor,
``Moduli potentials in type IIA compactifications with RR and NS flux,''
arXiv:hep-th/0411279.

\bibitem{Witten:1996bn}
E.~Witten,
``Non-Perturbative Superpotentials In String Theory,''
Nucl.\ Phys.\ B {\bf 474}, 343 (1996)
[arXiv:hep-th/9604030].

\bibitem{Katz:1996th}
S.~Katz and C.~Vafa,
``Geometric engineering of N = 1 quantum field theories,''
Nucl.\ Phys.\ B {\bf 497}, 196 (1997)
[arXiv:hep-th/9611090].

\bibitem{Berg:2004ek}
M.~Berg, M.~Haack and B.~Kors,
``Loop corrections to volume moduli and inflation in string theory,''
arXiv:hep-th/0404087.

\bibitem{Berg:2004sj}
M.~Berg, M.~Haack and B.~Kors,
``On the moduli dependence of nonperturbative superpotentials in brane inflation,''
%
arXiv:hep-th/0409282.


\bibitem{Kachru:2003aw}
S.~Kachru, R.~Kallosh, A.~Linde and S.~P.~Trivedi,
``De Sitter vacua in string theory,''
Phys.\ Rev.\ D {\bf 68}, 046005 (2003)
[arXiv:hep-th/0301240].

\bibitem{Maldacena:2000mw}
J.~M.~Maldacena and C.~Nunez,
``Supergravity description of field theories on curved manifolds and a no go theorem,''
%
Int.\ J.\ Mod.\ Phys.\ A {\bf 16}, 822 (2001)
[arXiv:hep-th/0007018].

\bibitem{deWit:1986xg}
B.~de Wit, D.~J.~Smit and N.~D.~Hari Dass,
``Residual Supersymmetry Of Compactified D = 10 Supergravity,''
Nucl.\ Phys.\ B {\bf 283}, 165 (1987).

\bibitem{Silverstein:2001xn}
E.~Silverstein,
``(A)dS backgrounds from asymmetric orientifolds,''
[arXiv:hep-th/0106209].

\bibitem{Maloney:2002rr}
A.~Maloney, E.~Silverstein and A.~Strominger,
``De Sitter space in noncritical string theory,''
[arXiv:hep-th/0205316].

\bibitem{Burgess:2003ic}
C.~P.~Burgess, R.~Kallosh and F.~Quevedo,
``de Sitter string vacua from supersymmetric D-terms,''
JHEP {\bf 0310}, 056 (2003)
[arXiv:hep-th/0309187].

\bibitem{Saltman:2004sn}
A.~Saltman and E.~Silverstein,
``The scaling of the no-scale potential and de Sitter model building,''
JHEP {\bf 0411}, 066 (2004)
[arXiv:hep-th/0402135].

\bibitem{Becker:2004gw}
M.~Becker, G.~Curio and A.~Krause,
``De Sitter vacua from heterotic M-theory,''
Nucl.\ Phys.\ B {\bf 693}, 223 (2004)
[arXiv:hep-th/0403027].

\bibitem{Buchbinder:2004im}
E.~I.~Buchbinder,
``Raising anti de Sitter vacua to de Sitter vacua in heterotic M-theory,''
Phys.\ Rev.\ D {\bf 70}, 066008 (2004)
[arXiv:hep-th/0406101].

\bibitem{Balasubramanian:2004uy}
V.~Balasubramanian and P.~Berglund,
``Stringy corrections to Kaehler potentials, SUSY breaking, and the cosmological constant problem,''
%
JHEP {\bf 0411}, 085 (2004)
[arXiv:hep-th/0408054].

\bibitem{Becker:2002nn}
K.~Becker, M.~Becker, M.~Haack and J.~Louis,
``Supersymmetry breaking and alpha'-corrections to flux induced  potentials,''
JHEP {\bf 0206}, 060 (2002)
[arXiv:hep-th/0204254].

\bibitem{Saltman:2004jh}
A.~Saltman and E.~Silverstein,
``A new handle on de Sitter compactifications,''
[arXiv:hep-th/0411271].

\bibitem{Frey:2003tf}
A.~R.~Frey,
``Warped strings: Self-dual flux and contemporary compactifications,''
[arXiv:hep-th/0308156].

\bibitem{Silverstein:2004id}
E.~Silverstein,
``TASI / PiTP / ISS lectures on moduli and microphysics,''
arXiv:hep-th/0405068.

\bibitem{Balasubramanian:2004wx}
V.~Balasubramanian,
``Accelerating universes and string theory,''
Class.\ Quant.\ Grav.\  {\bf 21}, S1337 (2004)
[arXiv:hep-th/0404075].

\bibitem{Kachru:2003sx}
S.~Kachru, R.~Kallosh, A.~Linde, J.~Maldacena, L.~McAllister and S.~P.~Trivedi,
``Towards inflation in string theory,''
JCAP {\bf 0310}, 013 (2003)
[arXiv:hep-th/0308055].

\bibitem{Blanco-Pillado:2004ns}
J.~J.~Blanco-Pillado {\it et al.},
``Racetrack inflation,''
JHEP {\bf 0411}, 063 (2004)
[arXiv:hep-th/0406230].

\bibitem{Denef:2004dm}
F.~Denef, M.~R.~Douglas and B.~Florea,
``Building a better racetrack,''
JHEP {\bf 0406}, 034 (2004)
[arXiv:hep-th/0404257].

\bibitem{Dasgupta:2004dw}
K.~Dasgupta, J.~P.~Hsu, R.~Kallosh, A.~Linde and M.~Zagermann,
``D3/D7 brane inflation and semilocal strings,''
JHEP {\bf 0408}, 030 (2004)
[arXiv:hep-th/0405247].

\bibitem{Burgess:2004kv}
C.~P.~Burgess, J.~M.~Cline, H.~Stoica and F.~Quevedo,
``Inflation in realistic D-brane models,''
JHEP {\bf 0409}, 033 (2004)
[arXiv:hep-th/0403119].

\bibitem{Kallosh:2003ux}
R.~Kallosh and A.~Linde,
``P-term, D-term and F-term inflation,''
JCAP {\bf 0310}, 008 (2003)
[arXiv:hep-th/0306058].

\bibitem{Firouzjahi:2003zy}
H.~Firouzjahi and S.~H.~H.~Tye,
``Closer towards inflation in string theory,''
Phys.\ Lett.\ B {\bf 584}, 147 (2004)
[arXiv:hep-th/0312020].

\bibitem{Hsu:2003cy}
J.~P.~Hsu, R.~Kallosh and S.~Prokushkin,
``On brane inflation with volume stabilization,''
JCAP {\bf 0312}, 009 (2003)
[arXiv:hep-th/0311077].

\bibitem{Hsu:2004hi}
J.~P.~Hsu and R.~Kallosh,
``Volume stabilization and the origin of the inflaton shift symmetry in string theory,''
%
JHEP {\bf 0404}, 042 (2004)
[arXiv:hep-th/0402047].

\bibitem{Buchel:2004qg}
A.~Buchel and A.~Ghodsi,
``Braneworld inflation,''
arXiv:hep-th/0404151.

\bibitem{Buchel:2003qj}
A.~Buchel and R.~Roiban,
``Inflation in warped geometries,''
Phys.\ Lett.\ B {\bf 590}, 284 (2004)
[arXiv:hep-th/0311154].

\bibitem{Kallosh:2004yh}
R.~Kallosh and A.~Linde,
``Landscape, the scale of SUSY breaking, and inflation,''
[arXiv:hep-th/0411011].

\bibitem{Shandera:2004aa}
S.~Shandera,
''Slow Roll in Brane Inflation''
[arXiv:hep-th/0412077].


\bibitem{Gukov:1999ya}
S.~Gukov, C.~Vafa and E.~Witten,
``CFT's from Calabi-Yau four-folds,''
Nucl.\ Phys.\ B {\bf 584}, 69 (2000)
[Erratum-ibid.\ B {\bf 608}, 477 (2001)]
[arXiv:hep-th/9906070].

\bibitem{Grimm:2004uq}
T.~W.~Grimm and J.~Louis,
``The effective action of N = 1 Calabi-Yau orientifolds,''
Nucl.\ Phys.\ B {\bf 699}, 387 (2004)
[arXiv:hep-th/0403067].

\bibitem{Klebanov:2000hb}
I.~R.~Klebanov and M.~J.~Strassler,
``Supergravity and a confining gauge theory: Duality cascades and chiSB-resolution of naked singularities,''
%
JHEP {\bf 0008}, 052 (2000)
[arXiv:hep-th/0007191].

\bibitem{Gross:1986iv}
D.~J.~Gross and E.~Witten,
``Superstring Modifications Of Einstein's Equations,''
Nucl.\ Phys.\ B {\bf 277}, 1 (1986).

\bibitem{Freeman:1986br}
M.~D.~Freeman and C.~N.~Pope,
``Beta Functions And Superstring Compactifications,''
Phys.\ Lett.\ B {\bf 174}, 48 (1986).

\bibitem{Grisaru:1986kw}
M.~T.~Grisaru, A.~E.~M.~van de Ven and D.~Zanon,
``Four Loop Divergences For The N=1 Supersymmetric Nonlinear Sigma Model In Two-Dimensions,''
%
Nucl.\ Phys.\ B {\bf 277}, 409 (1986).

\bibitem{Witten:1985bz}
E.~Witten,
``New Issues In Manifolds Of SU(3) Holonomy,''
Nucl.\ Phys.\ B {\bf 268}, 79 (1986).

\bibitem{Arkani-Hamed:2004fb}
N.~Arkani-Hamed and S.~Dimopoulos,
``Supersymmetric unification without low energy supersymmetry and signatures for fine-tuning at the LHC,''
%
arXiv:hep-th/0405159.

\bibitem{Arkani-Hamed:2004yi}
N.~Arkani-Hamed, S.~Dimopoulos, G.~F.~Giudice and A.~Romanino,
``Aspects of split supersymmetry,''
arXiv:hep-ph/0409232.


\end{thebibliography}
\end{document}